\documentclass[small]{acmart}

\AtBeginDocument{%
  }

\setcopyright{acmlicensed}
\copyrightyear{2018}
\acmYear{2018}
\acmDOI{XXXXXXX.XXXXXXX}
\acmConference[Conference acronym 'XX]{Make sure to enter the correct
  conference title from your rights confirmation email}{June 03--05,
  2018}{Woodstock, NY}
\acmISBN{978-1-4503-XXXX-X/2018/06}

\begin{document}

\title{Critical Challenges in Content Moderation for People Who Use Drugs (PWUD): Insights into Online Harm Reduction Practices from Moderators}

% Online Communities for People Who Use Drugs (PWUD): Critical Challenges for Content Moderators and Harm Reduction

% Critical Challenges in Content Moderation for People Who Use Drugs (PWUD): Insights into Online Harm Reduction Practices from Moderators

%Critical challenges content moderators face in supporting harm reduction for online PWUD communities

%Critical Challenges for supporting harm reduction for Online Communities for People Who Use Drugs (PWUD) through Content Moderation

\author{Kaixuan Wang}
\email{kw215@st-andrews.ac.uk}
\orcid{0000-0002-3795-0224}
\affiliation{%
  \institution{University of St. Andrews}
  \city{St. Andrews}
  \state{Scotland}
  \country{United Kingdom}
}

\author{Loraine Clarke}
\email{lec24@st-andrews.ac.uk}
\orcid{0000-0001-9213-1013}
\affiliation{%
  \institution{University of St. Andrews}
  \city{St. Andrews}
  \state{Scotland}
  \country{United Kingdom}
}

\author{Carl-Cyril J Dreue}
\email{v1cdreue@exseed.ed.ac.uk}
\orcid{0000-0002-4472-5746}
\affiliation{%
  \institution{University of Edinburgh}
  \city{Edinburgh}
  \state{Scotland}
  \country{United Kingdom}
}

\author{Guancheng Zhou}
\email{guancheng.joe.zhou@gmail.com}
\affiliation{%
  \institution{Independent Researcher}
  \city{Edinburgh}
  \state{Scotland}
  \country{United Kingdom}
}

\author{Jason T. Jacques}
\email{jtj2@st-andrews.ac.uk}
\orcid{0000-0003-3496-7060}
\affiliation{%
  \institution{University of St. Andrews}
  \city{St. Andrews}
  \state{Scotland}
  \country{United Kingdom}
}

\renewcommand{\shortauthors}{Wang et al.}


\begin{abstract}

Online communities serve as essential support channels for People Who Use Drugs (PWUD), providing access to peer support and harm reduction information. The moderation of these communities involves consequential decisions affecting member safety, yet existing sociotechnical systems provide insufficient support for moderators. Through interviews with experienced moderators from PWUD forums on Reddit, we examine the unique nature of this work and its implications for HCI and content moderation research. We demonstrate that this work constitutes a distinct form of public health intervention characterised by three challenges: (1) high-stakes risk evaluation requiring pharmacological expertise, (2) time-critical crisis intervention spanning platform content and external drug market surveillance, and (3) navigation of structural conflicts where platform policies designed to minimise legal liability directly oppose community harm reduction goals. Our findings extend existing HCI moderation frameworks by revealing how legal liability structures can systematically undermine expert moderators' protective work, with implications for other marginalised communities facing similar regulatory tensions, including abortion care and sex work contexts. We identify two necessary shifts in sociotechnical design: moving from binary classification to multi-dimensional approaches that externalise competing factors moderators must balance, and shifting from low-level rule programming to high-level example-based instruction. However, we surface unresolved tensions around volunteer labour sustainability and risks of incorporating automated systems in high-stakes health contexts, identifying open questions requiring HCI research attention. These findings inform the design of platforms that better accommodate vulnerable populations whose health needs conflict with regulatory frameworks.

\end{abstract}

% Contribution to Designing Interactive Systems:
% Our work reveals how platform governance frameworks and moderation systems impact harm reduction efforts within PWUD communities. We revealed opportunities to support their public health functions. Our findings contribute design implications for designing moderation tools that balance community safety with harm reduction needs while leveraging human expertise in high-stakes decision-making.



% %% A "teaser" image appears between the author and affiliation
% %% information and the body of the document, and typically spans the

\received{20 February 2007}
\received[revised]{12 March 2009}
\received[accepted]{5 June 2009}

\maketitle

\section{Introduction}
\label{sec: introduction}

Online platforms enable geographically dispersed individuals to form communities, creating spaces for connection and information exchange~\cite{marshall2024understanding}. For vulnerable populations, these communities are essential for accessing peer support and resources that are scarce offline. This is especially true for health-related communities, where users can discuss sensitive topics and share experiences with a degree of anonymity~\cite{schoenebeck2023online}. However, the open nature of these platforms exposes users to sociotechnical harms, which are the risks emerging from both platform design and social interactions~\cite{shelby2023sociotechnical}. These harms include exposure to misinformation threatening health outcomes~\cite{suarez2021prevalence}, harassment~\cite{schoenebeck2023online}, and content that may trigger harmful behaviours~\cite{marchant2017systematic, gillespie2018custodians}. 
Specifically for PWUD communities, these harms extend beyond typical platform concerns to encompass direct physical risks (e.g., dangerous dosage advice~\cite{soussan2014harm}) and systemic barriers (e.g., stigmatisation preventing help-seeking~\cite{paquette2018stigma}), necessitating content moderation.

As online communities have grown in scale and complexity, moderation practices have evolved from simple content review to sophisticated systems combining automated tools and human oversight~\cite{grimmelmann2015virtues}. While these systems aim to protect platform users from potential harms~\cite{schoenebeck2023online}, generic moderation approaches appropriate for general-interest communities (e.g., hobby and entertainment) may not always align with the needs of communities discussing high-stakes topics. In these spaces, moderation decisions can have direct consequences for a person's physical and mental well-being~\cite{chancellor2016thyghgapp, borras2021facebook, haimson2021disproportionate}.

Communities for People Who Use Drugs (PWUD) on social media exemplify the challenges of moderating high-stakes discussions online. For individuals who often face social stigma and barriers to offline medical care, these online spaces are a critical source of support~\cite{tighe2017information}. The central mission of these communities is harm reduction: a public health strategy and set of pragmatic approaches aimed at minimising the negative health and social consequences associated with substance use~\cite{marlatt2011harm}. In practice, this involves members sharing life-saving information on topics such as safer consumption techniques, overdose prevention, and identifying contaminated drug supplies \cite{davitadze2020harm, schwebel2023online}. However, conversations about safer consumption practices or advice on managing substance use disorders could be misinterpreted by automated systems as promoting drug use, potentially leading to the removal of harm reduction information~\cite{jain2020opioids, gomes2024problematizing}. Consequently, individuals seeking time-sensitive advice on overdose prevention may also face delays or lose access critical information due to overzealous content filtering.

To protect community safety and preserve open, supportive online spaces, moderators of PWUD forums must perform a difficult balancing act. They are responsible for enforcing community rules through moderation while ensuring members can still access vital harm reduction information. Moderating PWUD forums requires technical proficiency with moderation tools, a nuanced understanding of community norms~\cite{chancellor2018norms, costello2017online}, harm reduction expertise, and knowledge of the broader societal context of substance use \cite{rawat2022systematic}. This challenge is intensified by a structural conflict with platform and infrastructure providers, where platform-wide policies aimed at preventing the promotion of illicit activities may restrict the sharing of harm reduction information~\cite{gomes2024problematizing}. These rules can directly threaten the existence of PWUD communities on hosting platforms, forcing moderators to make difficult trade-offs between enabling open discussion and avoiding a community-wide ban for repeated policy violations~\cite{chandrasekharan2017you}, depriving PWUD of valuable support networks.

Though existing work has documented content moderation challenges faced in health support communities~\cite{marshall2024understanding, bizzotto2023loci, mccosker2023moderating, saha2020understanding, spence2024content, chung2022mental}, moderators' practices in PWUD communities and the challenges they face are not well understood. To address this gap, we examine the moderation practices in several PWUD-focused Reddit communities, including r/Drugs, r/MDMA, and r/researchchemicals. Reddit’s user-moderated structure provides an ideal context for studying how specialised governance practices develop in high-stakes environments~\cite{fiesler2024remember}. Through collaboration with moderator teams from these communities, including r/Drugs which served over 1 million subscribers as of September 2024, we explore how moderators navigate complex decisions around harm reduction content. Our study is guided by three broad questions:

\textit{RQ1: What are the distinct challenges and tensions in moderating online PWUD communities?}

\textit{RQ2: How do moderators develop and adapt practices to address these tensions while supporting harm reduction goals?}

\textit{RQ3: What opportunities exist for improving sociotechnical support for moderation in high-stakes harm reduction spaces?}

% This scope allows us to explore the unique difficulties these communities face, how moderators navigate decisions around harm reduction content, and where technological support could better serve their needs.

Our analysis reveals that moderating PWUD communities constitutes a distinct form of public health labour poorly supported by current sociotechnical systems. We make three contributions to HCI and content moderation research. First, we empirically identify three challenges defining this work: (1) high-stakes risk evaluation requiring pharmacological expertise to assess the safety of peer advice; (2) time-critical crisis intervention spanning both platform content and external drug market surveillance; and (3) navigation of structural conflicts where platform policies designed to minimise legal liability directly oppose community harm reduction goals. Second, we theoretically extend existing HCI frameworks on moderation trade-offs~\cite{jiang2023trade} by demonstrating how legal liability structures can systematically undermine expert moderators' protective work, revealing limitations in governance theories that assume aligned platform-community interests. Our findings inform design for marginalised online communities facing similar regulatory tensions, including abortion care and sex work contexts where platform policies conflict with community health needs~\cite{abortion2024, coombes2022disabled}. Third, regarding design, we identify two necessary shifts in sociotechnical support: moving from binary classification systems to multi-dimensional approaches that externalise competing factors moderators must balance, and shifting from low-level rule programming to high-level example-based instruction. However, we also surface unresolved tensions around volunteer labour sustainability and the risks of incorporating advanced technology, such as Large Language Models, in high-stakes health contexts, identifying open questions requiring HCI research attention. 

These findings have broader implications for designing moderation systems that support vulnerable populations where regulatory frameworks constrain life-saving information sharing. Global policy developments increasingly restrict online health spaces; for example, the UK's Online Safety Act\footnote{~\url{https://www.gov.uk/government/publications/online-safety-act-explainer/online-safety-act-explainer}} mandates identity verification, which undermines the anonymity that enables stigma-free support~\cite{onlinesafetyact2023}. In this context, understanding how moderators navigate these conflicts becomes essential for HCI researchers committed to designing platforms that serve marginalised communities' needs.

\section{Related Work}

Content moderation is the set of decisions and practices that determine what user content is allowed, restricted in visibility, or removed on online platforms~\cite{gillespie2018custodians}. As platforms have grown, moderation has moved from ad-hoc human review to sociotechnical systems that combine human judgement with automation~\cite{grimmelmann2015virtues, roberts2019behind}. Common approaches include rule-based filters (e.g., Reddit's AutoModerator\footnote{~\url{https://www.reddit.com/r/reddit.com/wiki/automoderator}}), which act on keywords or patterns specified by moderators specify~\cite{jhaver2019human, chandrasekharan2019crossmod}. Another approach uses machine learning classifiers trained on labelled data to detect problematic content at scale~\cite{gorwa2020algorithmic, pavlopoulos2017deep}. A recurring engineering premise in both approaches is that harmful content can be identified using stable criteria and handled consistently across contexts~\cite{muralikumar2023human}. This premise, however, often fails in settings where safety and health are at stake, as whether a post is helpful or harmful frequently depends on intent, situation, community norms, and platform policy. 

Online communities supporting People Who Use Drugs (PWUD) exemplify this tension. In these spaces, life-saving information about managing substance use risks can resemble content that platforms typically prohibit. This situation suggests broader challenges in moderating specialised health communities where standard content moderation frameworks, designed for general platforms, can systematically conflict with community-specific health needs. Understanding how moderators in PWUD communities navigate these conflicts is essential for designing governance systems that better serve vulnerable populations.

\subsection{Content Moderation and Governance on Social Media Platforms}

Content moderation serves multiple purposes on social platforms: protecting users from harm, facilitating productive interactions, enforcing community rules, and meeting legal obligations~\cite{gillespie2018custodians}. The balance among these functions varies substantially based on community norms and platform affordances. Recent research frames moderation as a form of community governance that requires negotiating among competing values and inherent trade-offs~\cite{jiang2023trade}. Platforms and community moderators must navigate persistent tensions between fostering open participation and preventing abuse, recognising that cooperative online spaces inherently entail some content restrictions~\cite{gillespie2017governance, vaccaro2021contestability}.

Empirical studies reveal that moderation strategies produce vastly different outcomes depending on the community context and implementation. A recent study documented that even a mild policy against ``incitement to hatred'' can trigger a backlash effect where hate speech increases following the policy's introduction~\cite{erickson2025content}. Such outcomes demonstrate how moderation interventions can backfire in communities resistant to oversight. Similarly, blanket bans on communities promoting hate speech, though aimed at reducing harmful content, have driven a substantial user exodus to alternative platforms~\cite{chandrasekharan2017you}. These platform-level interventions create displacement effects, potentially moving problematic behaviour to spaces with even less oversight~\cite{horta2021platform}. Beyond platform-imposed approaches, ``personalised moderation'' tools offer individuals control over what they see, which users value, yet simultaneously shift labour and responsibility from institutions to individuals~\cite{jhaver2023personalizing}. The tension between centralised regulation and distributed control reflects broader governance challenges for designing moderation systems, which the HCI community can help address in its fourth wave of research focusing on politics, values, and ethics~\cite{ashby2019fourth, sanchez2025let}.

\subsection{Content Moderation in High-Stakes Online Communities}
\label{sec:related_work1}

Unlike general-interest forums where moderation primarily affects user experience~\cite{jhaver2019human}, online communities dealing with \textit{high-stakes} contexts involve moderation decisions with direct consequences for users' physical safety or mental wellbeing~\cite{chancellor2016thyghgapp, chang2016managing, feuston2018beyond}. Examples include forums focused on mental health crises~\cite{saha2020understanding}, managing medical conditions~\cite{allison2021logging, delmonaco2023moderating}, addiction recovery~\cite{rubya2017video, garg2021detecting, chi2023investigating}, or other sensitive topics~\cite{perry2021suicidal, sorensen2023online}. Many such communities exist because traditional offline support, such as therapy sessions and in-person clinics, is inadequate or inaccessible due to cost~\cite{gaskin2012economic}, stigma~\cite{clement2015impact, park2023role}, or limited availability~\cite{schwarz2022barriers}.
These accessibility barriers drive individuals to depend on online spaces as critical alternatives for accessing timely advice, peer support, and health information that may not be obtainable elsewhere~\cite{sharma2018mental, wadden2021effect, lin2021social}.

The reliance on online communities for health information fundamentally reshapes what effective moderation must accomplish. Effective moderation practices in high-stakes communities demand specialised expertise~\cite{jiang2023trade}. Moderators must possess expertise beyond general content evaluation skills, needing deep understanding of specific medical terminology, mental health concepts, or substance interactions to accurately assess content quality and potential risks~\cite{seering2019moderator, kanthawala2021credibility}. Moderators typically develop this expertise not through formal training but through experience and active participation in the community~\cite{matias2019civic}. Hence, high-stakes communities often recruit moderators who not only understand platform policies but also possess lived experience and established trust within the community~\cite{atanasova2017exploring, seering2020reconsidering}. 
The collective knowledge in these communities frequently is codified into detailed guidelines reflecting community-specific considerations~\cite{chancellor2018norms}.

However, existing technical support constrains a moderator's ability to translate community expertise into consistent and effective actions (e.g., moderation policies or automated rules)~\cite{gorwa2020algorithmic}, which can lead to serious consequences in high-stakes settings. Chancellor et al.~\cite{chancellor2016thyghgapp} documented how Instagram's ban on specific hashtags related to pro-eating disorders led to community adaptation through lexical variants (e.g., \#thyghgapp instead of \#thinspo), resulting in content that became not only harder to detect but often contained more harmful messaging. Similarly, Feuston \& Piper~\cite{feuston2020conformity} found that content moderation of eating disorder discussions created pressure on users to conform or resist, often disrupting legitimate support-seeking for vulnerable individuals.

Moderation in high-stakes environments also faces time-critical pressures, where individual decisions may lead to immediate health risks for community members. Moderators must swiftly remove misinformation or misleading health advice to safeguard vulnerable users~\cite{huh2016personas, gillespie2018custodians, bizzotto2023loci}. Simultaneously, they must avoid overzealous moderation that censors important health information and cuts off access to life-saving resources~\cite{zhang2024debate, williams2023deplatforming}. Research demonstrates that applying generic rules without considering high-stakes contexts can unintentionally suppress valuable health insights, create barriers to support, or displace support-seeking into less visible channels~\cite{roberts2019behind, chandrasekharan2022quarantined, gillespie2022not}. Despite extensive research on moderation in various health communities, empirical studies examining these practices within PWUD communities remain limited.

\subsection{Harm Reduction and Online PWUD Communities}

Millions of individuals' wellbeing is affected by substance use, posing a challenge to the global public health landscape~\cite{unodc2023wdr}. Harm reduction has emerged as a vital strategy for minimising health risks and social harms associated with substance use by providing knowledge and support for PWUD~\cite{hedrich2021harm}. Online communities for PWUD host discussions about harm reduction practices and potentially life-saving information regarding safer drug use and overdose prevention~\cite{davitadze2020harm, schwebel2023online}, while offering stigma-free peer support for this vulnerable and marginalised group~\cite{bancroft2016concepts, van2016harm, tighe2017information}. 

However, platform regulations designed to prevent the facilitation of illicit activity often directly conflict with the vocabulary and topics required for harm reduction education~\cite{gomes2024problematizing}. These conflicts create ``collateral censorship,'' where policies targeting one form of harmful content inadvertently restrict legitimate health information~\cite{gillespie2018custodians}. When platforms implement broad content restrictions to manage legal liability, they inadvertently silence substance use discussions that constitute harm reduction education. Such policies reinforce the perception that only abstinence-based approaches to substance use are acceptable~\cite{piatkowski2024beyond, gomes2024problematizing}, limiting the spectrum of support available to PWUD online and potentially endangering lives when crucial harm reduction information becomes inaccessible.

Researchers have documented various moderation challenges in health-support communities, including difficulties in identifying health misinformation~\cite{malki2024headline}, managing the quality of information about mental health guidance~\cite{haime2025exploring}, and enforcing rules for platform users' mental health safety~\cite{gerrard2020covid19}. However, empirical studies that specifically examine challenges faced by moderators in PWUD spaces remain scarce. PWUD communities represent an analytically distinct case where legal constraints around substance use create systematic platform-community conflicts that undermine standard moderation approaches. This makes PWUD an important case for understanding the limits of current sociotechnical moderation frameworks when regulatory pressures directly oppose community health needs~\cite{gomes2024problematizing}. The absence of appropriate moderation frameworks for these communities can have severe consequences for individual well-being, potentially increasing overdose risks and preventing access to life-saving harm reduction strategies.

Reddit, as one of the world's largest social media platforms, provides a particularly rich context for studying these dynamics. Reddit's federated governance model delegates substantial authority to volunteer moderators, who translate platform policies into subreddit rules and day-to-day practice~\cite{fiesler2018reddit}. Its pseudonymous, topic-based organisation makes it especially suitable for examining sensitive health discussions when compared to identity-based networks like Facebook or broadcast-oriented platforms like Twitter~\cite{chan2025understanding, kahlow2024beyond}.
These affordances facilitate both privacy and community accountability simultaneously~\cite{park2023affordances, andalibi2016understanding}, making Reddit the primary peer support platform for PWUD communities and thus an appropriate environment to investigate moderation practices in sensitive contexts.

\subsection{Automated Tools Supporting Content Moderation}
\label{sec:related_work2}

Content moderation systems often combine an array of automated tools with human oversight to process high volumes of user-generated content~\cite{gillespie2018custodians, roberts2019behind}. 
Early interventions were simple, including outright content removal or basic keyword filters~\cite{srinivasan2019content}.
posts~\cite{zeng2022content} and adding warning labels to sensitive content~\cite{corradini2021investigating}, to balance community safety with information accessibility. 
The evolution of automated tools reflects ongoing research efforts to meet diverse community needs, progressing from basic rule-based systems to more sophisticated techniques. 

Reddit's automated moderation tool, AutoModerator (AutoMod), represents a rule-based approach. AutoMod allows community moderators to craft custom rules that automatically remove or flag content based on keywords or regexp patterns~\cite{jhaver2019human}.
This rule-based approach gives moderators control and can handle large volumes of posts, but it is inherently limited to what is explicitly encoded. Systems based on machine learning algorithms emerged to enable contextual content analysis beyond explicit keyword matching~\cite{pavlopoulos2017deep, gorwa2020algorithmic}. 
Such systems can detect subtler forms of policy violations (e.g., identifying hate speech~\cite{nobata2016abusive, fortuna2018survey}, harassment~\cite{gongane2022detection}, or other toxic content by analysing linguistic patterns) and have even been tailored to high-stakes scenarios like detecting signs of self-harm or suicidal intent in user posts~\cite{cohan2017triaging, chancellor2020methods}. 
There are also hybrid approaches: for instance, CrossMod is a system that uses machine learning to flag potentially harmful content across multiple communities, assisting Reddit moderators by sharing knowledge of trolls and ban-evasion across subreddits~\cite{chandrasekharan2019crossmod}. These advances demonstrate the potential of automation to scale up moderation and catch issues that might evade simple filters.

However, each approach to automated moderation involves distinct trade-offs that affect communities differently depending on the context. Rule-based filters like AutoMod are transparent and easy for moderators to tweak, but they can be brittle and lack nuance, potentially silencing needed discussion and pressuring conformity~\cite{feuston2020conformity}. For example, work on youth self-harm vernacular shows that expressions like ``I'm gonna KMS [kill myself]'' range from joking to imminent suicidal risk~\cite{ali2024m}. Over-flagging can suppress help-seeking, while under-flagging can miss time-critical crises, both with direct implications for member safety. 

Machine learning models can infer context beyond keywords, but they often struggle with domain-specific understandings. For example, widely used hate and abuse classifiers over-flag reclaimed slurs, negated statements (e.g., ``I don't hate...''), quotations, and counterspeech, while under-flagging more veiled abuse~\cite{rottger2021hatecheck}. These behaviours disproportionately silence members of marginalised groups who use reclaimed terms or informal language, while letting subtle attacks against them slip through. Additionally, the performance of generic machine learning classifiers transfers weakly across communities with biased formality and stereotypes rather than actual harm. This means informal or identity-affirming talk can be mislabelled as toxic, shaping moderation in ways that burden vulnerable users and distort support conversations~\cite{muralikumar2023human}. These findings suggest that existing algorithmic moderation can over-remove benign support from marginalised populations and miss targeted abuse against them, with clear risks to member safety and community integrity. More recently, Large Language Models (LLMs) present a greater ability to read longer contexts, infer user intent, and assist in moderators' decision-making processes. Yet, empirical evaluations report that LLMs inconsistently detect true violations and exhibit unstable behaviours under small changes in community norms, raising concerns about their implementation reliability and governance in practice~\cite{kolla2024llm, kumar2024watch}.

Given these limitations across all current automated approaches, human oversight remains crucial. Prior research emphasises that moderators still need to review algorithmic decisions and often must reconcile them with community norms and goals, for instance by overriding a platform's flag to keep an important discussion visible~\cite{seering2019moderator}.
Recognising these limitations, researchers have called for more human-centred and context-aware moderation tools~\cite{chancellor2023contextual}. Some have proposed adaptive moderation systems that better align with moderators' on-the-ground practices in sensitive contexts~\cite{lai2022human, haime2025exploring}. Rather than replacing moderators, such approaches integrate human insight with algorithmic support in a reflexive loop, allowing rules and models to be adjusted as community needs evolve~\cite{lai2022human, huang2025content}.

In summary, the existing literature shows a spectrum of automated moderation aids, from rigid filters to AI-driven classifiers, each with trade-offs in control, scalability, and accuracy. These lines of work inform our research by highlighting the gaps between what current tools offer and what high-stakes communities like PWUD actually need.

\section{Methods}
\label{Methods}

Online communities serve as valuable spaces for PWUD to seek harm reduction information and peer support. This study examined the challenges faced by moderators in PWUD communities, focusing on their practices, adopted tools, and the key trade-offs involved in maintaining these high-stakes spaces. We interviewed seven moderators from the three most prominent PWUD subreddit communities. Through collaboration with the moderator team on r/Drugs, researchers gained authorised access to moderation actions that are not publicly visible on Reddit. These included complete moderation logs\footnote{\url{https://support.reddithelp.com/hc/en-us/articles/15484543117460-Moderation-Log}} showing content removals, user bans, and other actions taken by both human moderators and Reddit's automated moderation system (AutoMod). AutoMod\footnote{\url{https://support.reddithelp.com/hc/en-us/articles/15484574206484-Automoderator}} operates through configurable rules that automatically take actions when posts or comments match specified patterns, keywords, or conditions. These rules can remove content, flag items for human review, or post automated responses with relevant resources. Access to both the logs and AutoMod configurations allowed us to analyse how moderators implement automated support. We used specific examples from this data during interviews to probe moderators' decision-making processes regarding moderation practices and rule implementation.

\subsection{Ethics Statement}
\label{sec:ethics}

This research was conducted with full ethical approval from our institution. We acknowledge the sensitive nature of discussions within PWUD communities and the broader societal debates surrounding substance use. To address this, we strove to maintain a neutral position, avoiding moral or ethical judgments regarding the moderators' work. Our approach aimed to understand the critical challenges in moderating PWUD communities and to provide qualitative insights into online harm reduction practices. Our findings may inform the design of governance frameworks and automated moderation tools that are more sensitive to the needs of PWUD groups. This research does not promote or endorse any specific practices related to substance use. Rather, it seeks to understand how these online communities are moderated to balance harm reduction practices, community safety, and open dialogue to better support marginalised communities.

\subsection{Community Scope}

We selected r/Drugs, r/researchchemicals, and r/MDMA as focal communities based on three criteria. First, we sought variation in community size and activity, with membership ranging from 220,000 to over 1 million, ensuring our findings would reflect moderation challenges across different scales. Second, all three communities explicitly centre harm reduction in their stated community guidelines. Third, we had established contact with r/Drugs moderators through prior research collaboration, which facilitated access and trust-building with the broader moderator team network.

These communities represent different facets of PWUD support on Reddit. r/Drugs serves as a general forum for discussing substance use across various contexts; r/researchchemicals focuses on emerging synthetic psychoactive substances requiring specialised expertise; and r/MDMA centres on the responsible use of MDMA\footnote{\url{https://en.wikipedia.org/wiki/MDMA}}. During recruitment, we discovered an overlap in moderator teams across these communities, with two individuals moderating multiple subreddits simultaneously. This overlap reflects the interconnected nature of PWUD moderation work on Reddit and strengthened our study by capturing moderators with broad expertise across different community contexts.

\subsection{Interviews}

We conducted semi-structured interviews with moderators from r/Drugs, r/researchchemicals, and r/MDMA. These PWUD subreddits, while overlapping in their moderator teams, vary in focus and size. Moderators were invited through Reddit messages. Their participation was voluntary, and we offered a 30~GBP Amazon gift card as appreciation for their time. The interviews were designed to understand the considerations involved in moderators' practices, such as the goals they pursued through moderation, the rationale for AutoMod configurations, and the trade-offs involved in making moderation decisions.

The interviews included 15 core semi-structured questions, supplemented with follow-up interviews. Most were conducted asynchronously via Reddit messages, with one synchronous Discord interview. Moderators were asked to express themselves as they preferred, including using informal language (e.g., slang) and emojis. This approach allowed us to enrich the interview data and capture social cues, which are often absent in text-only formats~\cite{hawkins2018practical}. A pilot interview was conducted with our collaborator, a moderator from r/Drugs, to refine our approach. 

Interviews were conducted within a similar timeframe, facilitating the cross-fertilization of findings~\cite{dahlin2021email}. During the interviews, we discussed specific examples of moderation decisions, including content not publicly visible, drawing on our authorised access to r/Drugs moderation decisions to provide context. This approach aimed to elicit insights into the trade-offs considered by moderators in their specific practices.

\subsection{Participant Recruitment}

These three subreddit communities are managed by a total of 18 moderators, with some volunteering on multiple subreddits. We invited 14 active moderators, defined as those who had performed any visible moderation actions in the preceding month, via Reddit direct message. 11 moderators initially consented to participate. However, some participants subsequently withdrew due to personal reasons or ceased responding to follow-up contact; we respected their decisions to withdraw from the study. The final analytical sample comprises seven moderators who completed the full interview protocol. 

These seven participants included three from r/Drugs, three from r/researchchemicals, and one from r/MDMA. Moderation experience ranged from three months to twelve years, with five participants moderating multiple PWUD subreddits simultaneously. Participants' backgrounds included individuals with harm reduction training, healthcare experience, a student pursuing a pharmacology degree, and those with lived experience in substance use communities. Collectively, they represented over 40\% of the active moderator population in these communities at the time of the study.

\subsection{Thematic Analysis}

The analysis began by preparing the data into a manageable corpus; we transcribed and formatted the interview data in Word documents. Table~\ref{tab:interview_data} summarises the data collected from each interview. To protect participant anonymity, each moderator is referred to by a participant number (P1, P2, P3, etc.) throughout this paper.

\begin{table}[h]
\centering
\caption{Interview data summary by participant.}
\label{tab:interview_data}
\begin{tabular}{lccc}
\hline
Participant & Interview Length & Word Count & Community (subreddit)\\
\hline
P1 & 11 pages & ~3,980 words & r/MDMA\\
P2 & 7 pages & ~2,950 words & r/Drugs \\
P3 & 9 pages & ~3,650 words & r/Drugs \\
P4 & 21 pages & ~4,150 words & r/Drugs \\
P5 & 10 pages & ~4,100 words & r/researchchemicals\\
P6 & 7 pages & ~2,450 words & r/researchchemicals\\
P7 & 5 pages & ~1,000 words & r/researchchemicals\\
\hline
\end{tabular}

\vspace{2pt}
\begin{flushleft}
{\footnotesize\textit{Note}: Some transcripts appear longer due to formatting, as moderators frequently referenced specific moderation actions and AutoMod configurations in their responses. The seven complete interviews generated approximately 22,280 words of transcript data.}
\end{flushleft}
\end{table}

Following Braun and Clarke's thematic analysis approach~\cite{braun2021thematic}, three researchers first familiarised themselves with the data, discussed notes, and exchanged initial insights. Transcripts were then uploaded to a dedicated NVivo project for collaborative analysis. The researchers coded the transcripts individually, ensuring that each interview was coded by two researchers. 
We employed ``open coding''~\cite{thornberg2014grounded} on a line-by-line basis. This method generates codes that stay close to the data and is valued for capturing the speaker's perspective as authentically as possible~\cite{bazeley2020qualitative}. This process resulted in 271 initial codes, including examples such as ``\textit{Communities aim for Harm Reduction}'' and ``\textit{AutoMod caused false positives to newcomers}.''

We conducted multiple iterative rounds of coding and note-taking. To ensure a shared understanding from multiple perspectives, we periodically discussed the codes, notes, and emerging concepts on a weekly basis throughout the analysis. We categorised the initial codes at a higher level after the first round. Codes that shared similarities were iteratively merged into more abstract concepts, resulting in 157 codes such as ``\textit{Experienced Tensions by Moderators}'' and ``\textit{Benefits of AutoMod}.'' Through subsequent rounds, we developed three key themes presented in the following subsections. After analysing data from five participants, we observed recurring codes, with subsequent interviews reinforcing the existing themes rather than introducing substantially new codes. The final two interviews confirmed the resulting thematic framework without introducing fundamentally new concepts, suggesting we had achieved adequate saturation for our study's scope~\cite{hennink2017code}.

\section{Findings}

We conducted interviews with moderators to explore the critical challenges they face on PWUD subreddits. Our analysis is guided by our research questions (see Sec.~\ref{sec: introduction}), which examine the unique tensions present in moderating PWUD communities, the practices moderators use to address them, and design opportunities for improving sociotechnical support in high-stakes harm reduction spaces. We begin by describing how moderators foster supportive communities as safe spaces for PWUD through content moderation in Sec.~\ref{sec: supportive_community}. Then, Sec.~\ref{sec: information_accessibility} details moderation practices and tools used to facilitate access to harm reduction information, ensuring it remains high-quality. Finally, Sec.~\ref{sec: protective_measures} presents the protective measures moderators engage in to safeguard the PWUD community from harm despite multiple constraints. 

\subsection{Fostering Supportive Communities for PWUD}
\label{sec: supportive_community}

Our first theme demonstrates the moderators' commitment to fostering a supportive community for PWUD by creating safe and non-judgmental spaces to share information and seek peer support. Their comments revealed a broader tension between the needs of PWUD groups and platform regulations. Moderators described their approaches to shaping the community toward open dialogue while managing emerging risks related to substance use. We detail these perspectives below, discussing first how moderators create safe and supportive spaces for PWUD, and second how they shape the community through moderation practices to support harm reduction goals, the core aim of their subreddits.

\subsubsection{Fostering Safe Spaces for Drug-Related Discussions and Support}

A key goal of moderation practices, stressed by all moderators, was creating safe, non-judgemental, and supportive spaces where PWUD can support each other and share harm reduction information involving controversial and legally sensitive topics. As P4 noted, ``\textit{Licit \& illicit drug use is part of our world}'', and the subreddits aim to ``\textit{minimize its [substance use's] harmful effects rather than simply ignore or condemn them}''. This approach counters the judgement PWUD frequently encounter in everyday life. P7 explained that their subreddit tries to achieve ``\textit{a community-wide focus on harm-reduction and evidence-backed information being promoted first and foremost.}'' Through practices rooted in harm reduction principles~\cite{frankeberger2023harm}, P2 commented that they hope to enable community members to ``\textit{help each other and improve the quality of life.}'' These quotes illustrate the moderators' central focus on establishing community spaces that acknowledge the unique need for PWUD to discuss controversial and legally sensitive topics, and sidestep judgmental attitudes to support harm reduction conversations. 

Moderators frequently emphasised the value of harm reduction subreddit communities for PWUD, as these spaces host ``\textit{basic advice that can save someone's life}'' (P2). P5 provided an example where ``\textit{some excessively potent neurotoxin,[substance names], made their way onto the market, and we were able to allow this to be known to the public}''. This demonstrates how moderation that provides access to open and non-judgemental drug-related discussions can prevent serious health consequences and warn community members of rapidly emerging health risks.

Moderators highlighted life-saving information is not always available on other platforms due to different governance strategies, making their subreddits crucial for PWUD as a safe space to gain harm reduction support. Moderators commented that PWUD experience censorship and stigmatisation on mainstream platforms. ``\textit{Most social media platforms straight up prohibit the types of discussions that happen here, making [the subreddit] a rare and much needed space}'' (P2). Moderators explained that access to safe harm reduction support for PWUD is limited by the governance frameworks adopted by other platforms.

To foster supportive and safe spaces for PWUD, moderators emphasised the importance of countering the stigmatisation PWUD face elsewhere. Moderators aim to provide ``\textit{non-coercive services and resources to PWUD}'' (P4), where individuals can access harm reduction information without being ``\textit{afraid of sharing their experiences}'' (P7). They consider their subreddits ``\textit{a place to share a passion that in all other aspects of our lives is hidden in the shadows}'' (P2). Moderators described how the creating non-judgmental spaces for discussing substance use enables PWUD to have open discussions and engage in peer support for better health outcomes that stigma often prevents.

In summary, moderators argued that the broader stigma and censorship against substance-related discussion on other mainstream platforms limit potentially life-saving support for PWUD in online spaces. Instead, they felt that Reddit hosts more open and accessible communities, allowing PWUD groups to support each other and share harm reduction information, such as life-saving advice on safer practice and urgent warnings about dangerous substances on the market. To foster these critical support networks, moderators emphasised the importance of and their commitment to creating a safe space for PWUD to help peers reduce health risks related to substance use and improve their quality of life.

\subsubsection{Shaping the Community to Support Harm Reduction Goals}

Beyond creating a safe space, moderators described actively shaping community norms around harm reduction by using moderation actions as educational opportunities and transparently communicating the rationale behind their decisions. They explained that their moderation practices are designed to build the community's understanding of harm reduction goals by disclosing the reasoning for their actions. Following a content removal, a message is sent to the poster of the content so that ``\textit{people on the receiving end of these decisions ... [can] have a better understanding of why I did what I did}'' (P3). Communicating the rationale for a moderation action serves multiple purposes. Another moderator highlighted their intention is to encourage ``\textit{the community member to learn that their behaviour is unwelcome and ... they should not repeat themselves [the behaviour]}'' (P6). 

Rather than ``\textit{simply issuing bans,}'' P2 described a goal to cultivate ``\textit{valuable community members}'' who contribute to what they call high-quality discussions and peer support. Furthermore, the transparency of moderation actions appears to strengthen community support for these practices. P5 observed that moderator posts generally ``\textit{are decently upvoted/well-received,}'' and P6 noted ``\textit{little to no negative opinion about our moderation practices.}'' This perceived positive reception suggests that such transparent practices do more than enforce rules; they build legitimacy and foster a sense of shared ownership over the community's harm reduction mission.

Several moderators further described how they adapt to address emerging substance-related risks. The moderation rules have been ``\textit{developed over 10 years as the changing moderators over the years have felt necessary}'' (P4). P2 explained that timely moderation intervention is needed when ``\textit{new problems arise}'':

\begin{quote}
``\textit{One thing that has been recently discussed with the mod team is the increase in posts about using diphenhydramine [DPH], which is ... associated with a lot of serious negative health impacts ... it is a drug easily sourced and abused by kids. The community consensus is that DPH use inherently goes against \textbf{the principles of harm reduction}.}'' - P2
\end{quote}

This quote demonstrates a shared understanding of harm reduction principles between moderators and community members, which guides decision-making to address emerging risks. Their collective assessment includes identification of health impacts, acknowledging realities facing vulnerable populations, and ensuring alignment with harm reduction goals.

Taken together, these practices reveal how moderators cultivate a community that actively supports harm reduction efforts. Through transparent, educational moderation, they build day-to-day trust that is critical when addressing emerging risks, as seen with DPH, where moderation becomes a tool to enact a ``community consensus''. Moderators foster a dynamic where the community collectively participates in the efforts required to keep each other safe.

\subsection{Facilitating Access to High-Quality Harm Reduction Information}
\label{sec: information_accessibility}

Our second theme describes how moderators facilitate access to harm reduction information and ensure it remains high-quality, up-to-date, and uncensored for PWUD. Moderators' experiences show how content curation and automated support guide the management of safety-critical information. Moderators described their efforts to preserve valuable discussions and remove content that introduces risks to others. They face technical and resource constraints in exercising their moderation practices effectively. Below, we detail their approach to facilitating access to high-quality harm reduction information, discussing how moderators curate community knowledge, utilise automated support, and navigate technical constraints in maintaining access to harm reduction resources.

\subsubsection{Curating Community Knowledge While Minimising Substance-Related Risks}

Moderators described their efforts to curate high-quality harm reduction information from community discussions, noting that this directly impacts individual safety. To increase the availability of high-quality harm reduction information and minimise deleterious content, moderators use both restrictive and permissive approaches. Moderators reported that they restrict posts lacking detailed descriptions or containing easily searchable questions, as ``\textit{simple questions can result in more complex and useful discussions getting drowned out}'' (P4). This includes inquires about first-time experiences with substances that are well-documented in the existing harm reduction resources. Several moderators highlighted that their community guidelines explicitly require users to search their questions on relevant websites first. This prevents ``\textit{burdening subscribers}'' (P3) with duplicate content and ensures discussions extend meaningfully to the community's knowledge base. These quotes demonstrate that moderators curate community knowledge through restrictive approaches, ensuring community discussion focuses on substantive harm reduction topics. 

Moderators also described permissive approaches to preserve valuable harm reduction discussions. When detailed posts are not feasible, moderators explained that they provide specific guidance to help users bypass what P4 calls the ``\textit{junk filter},'' such as explicitly stating their prior search efforts. More significantly, P5 explained that they sometimes preserve rule-violating content when its educational value outweighs potential risks. They went on to explain this reasoning through a specific example:

\begin{quote}
``\textit{I used to remove any post ... mentioning that they take say, opioids and benzodiazepines, or with maybe alcohol ...[but] maybe it would help enforce to the poster that it is indeed a bad idea if they have multiple people commenting that it is so on their thread, vs just one single post-removal comment from the moderators about it.}'' - P5
\end{quote}

Rather than simply removing discussions about dangerous substance combinations, moderators reported that they adopted permissive approaches enabling community members to reinforce harm reduction principles collectively. These permissive designs reflect moderators' recognition that peer responses can strengthen educational impact in ways that moderator interventions alone cannot.

Similarly, moderators noted nuanced curation strategies when handling rule violations. P1 described adjusting ban severity based on how it might affect harm reduction discussions: ``\textit{if we determine that it would be detrimental to the users' ability to interact with our userbase about harm reduction, then we give somewhat less severe bans so that we don't limit harm reduction information for said user.}'' This case-by-case approach extends to repeat violations. P6 explained their graduated response:

\begin{quote}
``\textit{I usually provide some leeway for users who have broken the rules once or a couple of times. ... However, repetitive breaking of the rules [by a user] will result in a temporary ban... or permanent ban [if more violations happen]...}'' - P6
\end{quote}

The considerations involved in these moderation decisions are further illustrated through moderators' responses to specific cases. When presented with a post that had been removed by AutoMod rules for its aggressive tone despite promoting responsible substance use education, P3 expressed that they would reverse this automated decision as its educational value should be prioritised: ``\textit{The way I see this post is as a great opportunity for discussion, which aligns with what [the subreddit] is about.}'' These examples demonstrate how moderators flexibly evaluate rule violations to facilitate what they see as ``high-quality'' knowledge exchange. This approach has the potential to spark novel, peer-driven educational discussions that reduce harm and are not available elsewhere.

In summary, moderators employ both restrictive and permissive approaches to manage the flow of critical and potentially life-saving information. For PWUD, valuable knowledge often emerges from lived experience. Moderators prioritise a post's educational potential over simple rule enforcement, facilitating access to harm reduction discussion emerging from community interactions.

\subsubsection{Implementing Automated Support for Information Management}
\label{sec: automated support}

Moderators view AutoMod as a valuable tool to help improve information quality on their subreddits and facilitate access to harm reduction information for PWUD. AutoMod can effectively remove irrelevant discussions, such as promoting products and ``political beliefs'' like abstinence, ensuring the community ``\textit{stays relevant}'' (P5) and focused on harm reduction. AutoMod's swift enforcement prevents what P4 described as ``\textit{a self-reinforcing feedback loop of low quality content}'' that can emerge when rule violations remain visible. In a high-stakes environment where urgent queries about substance safety compete for attention, such a loop could bury critical, time-sensitive information and undermine their subreddit's core purpose of providing harm reduction resources. P3 explained that by freeing moderators from tedious work like removing unwanted content, AutoMod allows them to focus on ``\textit{keeping the community engaged and up to date on news}''.

To ensure users' access to community support, moderators stated that they include relevant resources in the automated responses sent after moderation actions. When users ask for information that does not align with the focus of their subreddit, an automated response is sent to ``\textit{direct them to another community where their post may be more suitable}'' (P6) or to advise on edits ``\textit{so that their post does not breach any rules}'' (P6). P4 highlighted that they use automated replies to guide users to reliable resources like Erowid\footnote{\url{https://www.erowid.org}} and TripSit\footnote{~https://tripsit.me}, to enhance the ``\textit{availability and visibility}'' of harm reduction information. Moderators use automated tools to transform content filtering into pathways for sharing harm reduction information. The integration of external resources into automated responses demonstrates how technical implementation can be leveraged to provide PWUD with access to critical resources.

Moderators' use of automated support extends to managing life-threatening situations. For example, they identified a frequent overlap between discussions of substance use and expressions of suicidal ideation. P2 commented that suicide-related support communities represent some of the ``\textit{most common subreddits people are referred to.}'' When community members show signs of crisis, moderators employ a multifaceted response to help individuals. P5 described how moderators first report concerning posts to Reddit administrators, and then use automated system messages to connect individuals with specialised support communities including ``\textit{r/Addiction, r/Depression, r/MMFB (Make Me Feel Better), r/SuicideWatch, and r/StopSpeeding.}'' These responses show how moderators utilise automated mechanisms to address the interconnected mental health crises they frequently encounter and ensure harm reduction information reaches the original posters. 

\subsubsection{Technical and Resource Constraints in Moderating Substance-Related Content}
\label{sec: constraints}

While automated tools are vital for managing the high volume and quality of content and directing users to appropriate resources, their use in harm reduction contexts presents challenges. Moderators reported facing technical limitations with these automated tools, alongside resource constraints. These factors create ongoing tensions between the need for rapid, scalable moderation and the need for context-sensitive actions to deliver relevant safety information to community members. We further explain the technical and resource constraints they described in the following sections. 

\paragraph{Technical Constraints.} Moderators reported multiple difficulties with AutoMod's configuration such as incorrect content removals and the challenge of maintaining consistency across moderation teams. These issues particularly affect moderators who inherit rather than develop automated rules. As P3 explained: ``\textit{I'm not the person that has written these rules it's sometimes hard for me too to completely...understand/make connections between specific word/word combinations used in a post/comment and if it breaks a subreddit rule or is unwanted for other reasons.}''

As communities grow and more rules are configured into AutoMod, some moderators find it difficult to maintain up-to-date configurations. P4 described how legacy rules designed for specific threats can persist beyond their usefulness, potentially leading to incorrect automated decisions: ``\textit{We had a spam campaign for some time where the scammers often used the spelling mistake 'stuffs'. There are still rules on my subreddit that look for this, but the scam campaign has stopped.}'' The challenge extends beyond outdated rules to the broader difficulty of maintaining comprehensive knowledge of automated systems. As P4 notes, ``\textit{the main problem is that I cannot memorise all the rules.}'' These examples show how technical constraints emerge from the accumulation of automated rules, as moderators struggle to maintain and interpret the configurations they inherit. The persistence of outdated rules and difficulties in understanding rule interactions illustrate the technical limitations of current automated tools, which often impede the effective management of harm reduction content and block PWUD's access to safety-critical information.

Moderators expressed a need for tools more effective than AutoMod to manage the high-stakes discussions on PWUD subreddits. While P4 acknowledged AutoMod as ``\textit{the least-imperfect moderation tool available,}'' moderators consistently highlighted its lack of context sensitivity as a critical weakness that diminishes the community efforts to promote harm reduction information. P2 noted frequent `false positives' in content removal that directly impede PWUD's access to community support. P4 described one such challenge with automated rule enforcement:

\begin{quote}
``\textit{lots of anti-sourcing rules hope to target present and future tense actions, but often do not need to target past tense descriptions. So 'where can I get heroin on vacation?' should be removed but 'I went on vacation and got high on heroin, here is my report' would not need to be removed.}'' - P4
\end{quote}

They went on to give another example of an incorrect removal made by AutoMod because the discussion shared a similar semantic meaning to ``drug test'', which is forbidden on their subreddit. We anonymised the personally identifiable information here:

\label{exp:automod-removal-blood-test}
\begin{quote}
``(A post was removed by AutoMod, shared by P4)

\textit{Question: How long does [a substance] last in blood?}

\textit{Description: I understand the information online however I'd like to see the community's response to this. [Description of poster's substance use] Anyone have any experience with blood tests for the window detection period? [Asking for advice on cleansing the drug in the body] Thanks everyone}''
\end{quote}

This inaccurate automated removal demonstrates how technical limitations in moderation systems can impede PWUD's access to safety-critical information and peer support. P6 explained that other types of valuable information are at risk of being removed by AutoMod, including research or experience reports on novel substances, as they can ``\textit{prevent some(body) from harming themselves,}''. However, P4 noted that ``\textit{[configuring AutoMod to perform accurately] is not easy.... as users may deliberately circumvent these filters by changing their typing, such as changing `I need a dealer' to `I need a D e a l e r'.}''

These situations reveal how current automated tools' limited ability to interpret context creates ongoing challenges for moderators trying to preserve harm reduction discussions that can reduce future harm. Moderators' comments also suggest that current automated moderation tools cannot accurately distinguish between posts enabling harm (e.g., sourcing) and those seeking to prevent it (e.g., sharing lived experiences), which undermines the subreddit's ability to circulate life-saving, peer-generated knowledge.

\paragraph{Resource Constraints.} While some moderators reported sufficient labour to provide effective moderation for their respective subreddits, many face labour shortages that affect their ability to review flagged content and maintain community standards. P4 stressed that increasing discussion volume strains moderation resources, particularly given the demanding requirements for new moderators to be: ``\textit{`active 4+ days per week'} and possess ``\textit{some drugs and harm reduction expertise, so that they can understand the context of things.}'' The demanding combination of time commitment and harm reduction expertise suggests an inherent tension between moderating at scale and maintaining context-sensitive support for PWUD communities.

\paragraph{Design Opportunities.} When discussing how new tools might enhance PWUD community moderation, we raised suggestions outlining how Large Language Models (LLMs) could generate personalised harm reduction information and analyse discussion topics. Moderators believed new tools such as LLMs could improve moderation accuracy and help navigate harm reduction information more effectively. P4 envisioned that leveraging LLMs ``\textit{could vastly improve [current moderation practices] by providing more relevant information and less irrelevant information [to community members], hopefully leading to people actually reading what we send them.}'' 

P1 proposed targeted applications that can be enhanced by adopting LLMs, focusing on accurate content prioritisation and moderation workflows:

\begin{quote}
``\textit{Having a way to make a priority system for content removal/flagging would be nice... it could have a high priority to posts that are obviously scam/spam [and] flag for manual reviews [for] posts pertaining to drug interactions. Marking comments that might be promoting harm ... would also be useful.''}
\end{quote}

Moderators' responses reveal how they envision automated tools not simply as replacements for manual work, but as a means to enhance their judgment in critical content decisions. Their focus on content prioritisation and targeted review mechanisms demonstrates how technical implementation could support rather than supplant human expertise in harm reduction contexts.

In summary, moderators' comments reveal that accumulated AutoMod configurations, limited context sensitivity, and the specialised expertise required for moderating harm reduction PWUD communities are ongoing hurdles to leveraging current automated moderation tools. Technical limitations and resource constraints simultaneously impact moderators' ability to implement automated tools effectively and their efforts to facilitate access to time-sensitive, and potentially life-saving, harm reduction discussions. This suggests opportunities for better tools that support human judgment in moderating PWUD communities.

\subsection{Protective Measures in Moderating PWUD Communities}
\label{sec: protective_measures}

The third theme discusses the protective measures moderators adopt to sustain PWUD online communities and safeguard individual members. When discussing community protection, moderators described tensions between complying with platform policies and maintaining access to harm reduction information. Their efforts to protect individual members demonstrate their reliance on specialised harm reduction expertise to combat potentially dangerous content. We detail these findings below, examining first how moderators preserve community existence within platform constraints, and second how they leverage their expertise to safeguard individual members.

\subsubsection{Preserving Community Existence Within Platform Constraints}
\label{sec: preserve_community}

PWUD community moderators emphasised that they prioritise preserving the community's existence by strictly complying with Reddit's ban on substance sourcing discussions \footnote{\url{https://support.reddithelp.com/hc/en-us/articles/360043513471-Reddit-s-policy-against-transactions-involving-prohibited-goods-or-services}}. P5 commented that ``\textit{Reddit has made their views on sourcing very clear, and has taken down plenty of subreddits for violation}.'' Moderators discussed the implications and tensions they face in trying to provide harm reduction whilst avoiding being shut down. For the subreddit to exist, they resort to strict enforcement, as P7 explained: ``\textit{upholding legality rules is a must if we plan to keep this community on the face of a public forum like Reddit.}'' As a result, the subreddit's community guidelines explicitly state:

\begin{quote}
``\textit{Strictly no requesting, mentioning or giving sources of drugs or paraphernalia, whether legal or illegal. If in doubt, then DON'T. If your post, or a reply to it would make it easier for someone to get drugs, it's not permitted. This includes sourcing conducted in private messages.}'' - P4
\end{quote}

However, even though the rule is highly visible on the community's front page, several moderators reported that drug sourcing is one of the most frequent rule violations. P5 stated that ``\textit{by far, sourcing-violation posts are the most frequently encountered.}'' As such, they stressed the need to adopt a zero-tolerance policy for drug sourcing conversations when moderating the subreddit. P1 noted, ``\textit{For some rule breaking (like the sourcing rule) we either straight out permanent [sic] ban or give lengthy bans}''. These responses reveal how the moderators' commitment to preserving the community's existence compels them to enforce strict rules, reflecting a unique tension in moderating PWUD communities.

Moderators consider AutoMod to be a valuable tool to help them enforce these protective measures, as it can be configured ``\textit{with names of different vendors ... [which] does effectively keep a lot of sourcing discussions off the page}'' (P5). However, such strict enforcement using AutoMod has led to incorrect removals of genuine harm reduction content; moderators view this as a necessary trade-off. ``\textit{It's a compromise I don't like, but better safe [to keep community existence] than sorry}'' (P3). P5 further explained how platform policies and current restrictions limit proven harm reduction practices:

\begin{quote}
``\textit{if people were able to review other vendors, like Reddit pre-2017 (when they enforced no-sourcing officially), that better harm reduction would be taking place since it would be more known when bad batches, low-quality, scams, etc are taking place.}'' - P5
\end{quote}

These responses show how changes in platform policies affect PWUD communities' ability to host harm reduction information. By prohibiting contextually life-saving information, these policies force moderators to restrict content to preserve their existence on Reddit.

In an attempt to balance these restrictions with harm reduction goals, Reddit permits discussions about sourcing harm reduction paraphernalia, ``\textit{such as scales, testing kits, syringes, micron filters and so on}'' (P4). 
This flexibility in platform policy demonstrates how governance frameworks could better support harm reduction efforts through nuanced distinctions rather than blanket prohibitions.

In summary, moderators' insights indicate that the platform policies have pushed PWUD subreddits to restrict discussions that could reduce risks related to substance use in order to preserve their existence on Reddit. Their efforts to protect these vital online spaces through strict content control reflect the high cost of compliance with evolving governance frameworks. The permissibility of safety equipment discussions shows how platform rules could evolve to better support harm reduction needs.

\subsubsection{Safeguarding Individual Members Through Moderation}

Beyond community-level protection, moderators reported strategies to safeguard individual members from potential harms associated with misinformation or dangerous advice. Moderators noted a keen awareness of the complexities surrounding harm reduction advice in online spaces. They recognised that while peer advice can be invaluable, it also carries inherent risks, particularly when dealing with substances and practices that can have severe health consequences. As P1 noted, ``\textit{The responses our users give are subject to moderation, because not everyone is well versed in the pharmacology of drugs, even if they are genuinely trying to help}''. This illustrates their reliance on pharmacological knowledge to moderate content and leverage such expertise to protect members.

Several moderators highlighted their approach to safeguarding individual members from misinformation by restricting substance identification requests. P3 explained that ``\textit{most people can't answer [queries asking to identify a specific substance] because it's a niche very little [sic] people know the details about.}'' For example, the same substance can exist in dramatically different physical appearances. Instead, moderators encourage community members ``\textit{to use things such as reagent testing, or analytical laboratory services}'' (P5) as moderators aim to educate PWUD about practices that reduce the risks related to substance use. Similarly, moderators restrict discussions about drug pricing to ``\textit{prevent our users from getting direct messages from malicious users who want to scam desperate people searching for drugs}'' (P1). In response, moderators constantly tune keyword filters to ``\textit{stop such spam attacks [such as automatic bots posting content promoting cryptocurrencies or malicious intent] on the sub.}'' (P1) 

These protective measures, aiming to address specific risks facing individual PWUD members, demonstrate moderators' efforts to safeguard individuals in practice. Their prioritisation of scientific testing over crowd-sourced substance identification reveals their intention to minimise the risks associated with peer advice and protect community members from potential harm.

Another safeguarding activity moderators use to protect individuals' safety is removing egregiously dangerous advice. P5 discussed a specific example of a rule violation where dangerous advice was presented as being safe:

\begin{quote}
``\textit{...we try to exist for education, and harm reduction. Sometimes there is just a bad apple that will be there to troll users with horrible advice. Say someone is looking to take oDSMT, an opioid analogue of the prescription Tramadol, has no tolerance to opioids, and is curious what dosage would be best... Say someone comments 'Try 200mg, that's a great dose!....' This would be a very excessive dose for a beginner anyways ... potentiate this into a potentially lethal situation.}''
\end{quote}

P6 described other types of harmful advice, listing ``\textit{unsafe substance use and downplaying the negative and incredibly harmful, effects of chronic use, combining drugs, and high dosage use, despite giving anecdotal experience reports.}'' In response to these situations, P1 reported that they issue bans to anyone posting irresponsible drug use content or malicious questions to protect individual members' safety from the inherent risks linked to peer advice.

These examples show how moderators need and leverage extensive harm reduction expertise to identify and address varying levels of risk within the community discussions. Their detailed understanding of substance interactions and dosage thresholds enables critical assessment of potentially dangerous content, ranging from reports of excessive substance use to advice recommending lethal substance combinations. 

In summary, moderators' protective measures involve complex safeguarding at multiple levels to protect their PWUD communities. At the community level, they adhere to platform policies banning substance sourcing while finding ways to preserve critical safety information. At the individual level, their pharmacological expertise is critical in evaluating potentially harmful and dangerous substance-related advice.

% In summary, moderators' protective measures involve complex safeguarding at multiple levels to protect their PWUD communities. At the community level, they adhere to platform policies banning substance sourcing while finding ways to preserve critical safety information. At the individual level, their pharmacological expertise is critical in evaluating potentially harmful and dangerous substance-related advice. These practices reveal the specialists knowledge required spanning x y z (insert area of expertise needed) to protect both community resources and individual members and the strategies they adopt to do this. 

%These practices reveal how moderators' specialized knowledge together shape their ability to protect both community resources and individual members.

\section{Discussion}

Substance use affects approximately 300 million individuals globally, accounting for over 3 million deaths annually~\cite{unodc2024wdr, who2024death}, many of which are preventable with timely harm reduction intervention. However, stigma pervades healthcare settings, where PWUD face discrimination from medical professionals, leading to delayed care-seeking and deteriorating health outcomes~\cite{paquette2018stigma, papalamprakopoulou2025breaking}. Online communities have emerged as alternative spaces where PWUD exchange harm reduction information to manage substance use risks, improve health outcomes, and potentially save lives~\cite{tighe2017information}. Yet our study reveals that major platforms hosting these communities provide insufficient support for the moderation of these spaces. Policies designed to minimise legal exposure systematically constrain harm reduction practices~\cite{gomes2024problematizing}, and this restrictive a trend is escalating. The UK's Online Safety Act\footnote{\url{https://www.gov.uk/government/publications/online-safety-act-explainer/online-safety-act-explainer}}, enforced from July 2025, mandates identification verification for users accessing platforms with adult or potentially harmful content~\cite{onlinesafetyact2023}. This creates new barriers for moderators cultivating supportive spaces for open harm reduction education. For PWUD discussing illegal substances, identity verification creates legal risks that undermine the anonymity enabling these communities to function as stigma-free support spaces. These policy developments also threaten research access. Researchers may be unable to conduct the studies needed to understand this uniquely challenging labour and develop appropriate sociotechnical systems.

The challenge, which is rooted in conflicts between regulation targeting illicit activities and communities' health goals, aligns with recent calls within HCI to foreground politics, values, and ethics in design, pushing beyond critique towards institutional change~\cite{sanchez2025let}. How can HCI progress towards values-centred design for vulnerable populations without understanding the communities we aim to support, especially as policy changes increasingly restrict that access? This study takes a step toward addressing this question by revealing the specific moderation challenges in online PWUD communities. It highlights labour that combines high-stakes risk evaluation, time-critical crisis intervention, and the navigation of platform-community conflicts. Understanding these challenges matters for both the HCI community and harm reduction researchers. Our findings offer insights for designing systems that support other marginalised populations facing similar regulatory tensions, including abortion care communities where platform policies restrict health information sharing~\cite{abortion2024} and sex workers who experience systematic deplatforming despite adverse impacts on their safety~\cite{coombes2022disabled}. The discussion below examines each identified challenge and proposes design directions grounded in our findings.

\subsection{Challenges of Moderating Online PWUD Communities}

High-stakes health communities face documented moderation challenges where content decisions directly affect members' physical safety and mental well-being~\cite{feuston2020conformity, perry2021suicidal, delmonaco2024you}. These moderation actions can have disproportionate negative implications for marginalised groups over gender and race~\cite{haimson2021disproportionate}. PWUD represent a community that both shares high-stakes contexts where ``basic advice that can save someone's life'' (P2) and experiences marginalisation due to societal stigma. In other high-stakes areas like mental health support~\cite{ahmed2022machine} and suicide prevention~\cite{abdulsalam2024suicidal}, the goals of platforms and communities are generally aligned, allowing for the creation of datasets where expert-labelled ``harmful content'' is less controversial. For PWUD, however, the tension between platform regulation and community-based harm reduction practices complicates the very definition of what is harmful or helpful~\cite{gomes2024problematizing}.
Reddit specifically, a platform viewed as much more open by moderators than other mainstream platforms, prohibits discussions facilitating the acquisition of controlled substances to limit legal liability. From a harm reduction perspective, some prohibited discussions serve protective functions. Before Reddit's 2017 ban on substance sourcing, community members posted vendor reviews warning about ``bad batches, low-quality, scams'' (P5). When contaminated supplies appeared, communities rapidly identified and publicised the risks. However, the policy eliminated this pathway. Moderators must now remove vendor discussions to prevent community bans, even when those discussions would prevent overdoses or poisonings. 

This misalignment changes what moderation work entails. Moderators must balance regulatory compliance against community safety, accepting a ``better safe than sorry'' (P3) approach even when AutoMod incorrectly removes legitimate content. They adjust ban severity based on whether enforcement ``would be detrimental to the users' ability to interact with our userbase about harm reduction'' (P1), weighing community survival against individual information access~\cite{jiang2023trade}. Decisions operate under a constant question: will this specific action trigger platform enforcement? This practice differs from evaluating norm violations; instead, moderators manage institutional risk alongside content curation. Moderation also extends beyond platform content. Unregulated drug markets evolve rapidly, with new compounds appearing without warning. This creates time-critical threats that formal systems cannot track~\cite{moon2024enhancing}. P5 described monitoring for ``excessively potent neurotoxin'' substances and ensuring communities could warn members. Thus, moderators track substance evolution, policy changes, and emerging risks, performing continuous surveillance both on and off the platform.

However, moderators' efforts to promote harm reduction education and protect community members' safety can be significantly undermined by the platform's regulatory requirements. This dynamic reflects a privilege that can marginalise vulnerable communities' expertise~\cite{ajmani2024whose}. As Gillespie observes, ``the margin of error typically lands on the marginal''~\cite{gillespie2020content}, where overzealous moderation disproportionately affects already-marginalised groups~\cite{haimson2021disproportionate}. PWUD moderators restrict health information to preserve the community's existence, enacting the very deprivation their expertise exists to prevent. Current HCI frameworks often assume that content is harmful for all platform users and should be removed. Yet, they offer no guidance when harm emerges from enforcement itself, when health goals oppose regulatory requirements, or when effective expertise cannot be exercised~\cite{jhaver2024bystanders, wilms2025technology}. This gap demands that we reconceptualise how platforms accommodate communities whose health needs conflict with legal liability frameworks.

\subsection{Socio-Technical Support for Moderators in Online PWUD Communities}
Addressing the challenges faced by moderators in online PWUD communities requires socio-technical support built through collaboration among stakeholders. Moderators reported that automated systems like AutoMod, which operate on syntactic pattern-matching, are insufficient for the nuances of harm reduction discourse. These tools struggle to interpret the contextual cues that differentiate prohibited content from valuable safety information, as shown by the incorrect removal of a member's help-seeking post about a blood test due to an efficient but simple keyword match (see Sec.~\ref{exp:automod-removal-blood-test}). While the inadequacy of rule-based moderation is a recognised problem that can lead to the removal of legitimate health content~\cite{haime2025exploring}, with serious implications for individuals' well-being, simply adopting existing solutions, such as developing AI tools trained on annotated data~\cite{sarridis2022leveraging, gongane2022detection}, can be problematic for PWUD communities. We argue that PWUD moderators' work can benefit from support that shifts the focus from simply deciding \textit{what to remove} towards \textit{externalising the trade-offs} discussed earlier, keeping the moderation decisions contestable~\cite{vaccaro2021contestability}.

Under the current regulatory landscape, one direction forward involves annotation schemas capturing the competing factors moderators navigate: poster intent (e.g., [Intent: Seeking Health Information]), harm reduction value (e.g., [Benefit: Drug Safety Query]), and platform policy risk (e.g., [Risk: Sourcing Keywords Present]). This makes trade-offs explicit, supporting moderator's decision-making process. Developing such schemas requires collaboration between harm reduction practitioners, platform developers, and active moderators who understand community norms and regulatory constraints. However, critical questions remain unresolved. Who provides annotations? If moderators annotate while moderating, the labour burden increases substantially; P4 described the demanding requirements of being ``active 4 days per week'' plus possessing ``drugs and harm reduction expertise.'' Volunteer moderation at this intensity is unsustainable~\cite{kiene2019volunteer}. 

Recent HCI work estimates Reddit's volunteer moderation represents at minimum \$3.4 million in annual uncompensated effort~\cite{li2022measuring}, with moderators reporting burnout from managing growing communities~\cite{dosono2019moderation}. External annotators lack the community-specific knowledge needed to assess harm reduction value accurately. Platforms face potential legal liability if courts determine specialised moderators should be compensated~\cite{matias2019civic}. This tension, between designing systems that support expert judgment and not overburdening the volunteers performing that judgment, remains unresolved. Addressing this requires HCI research on low-burden annotation approaches, potentially through retrospective labelling of historical content for model training rather than imposing real-time moderation burdens, or examining whether platforms should directly compensate moderators performing specialised high-stakes work. 

The time-critical nature of moderation in PWUD communities creates another requirement to effectively triage high-risk content, imposing operational burdens due to current tool design. Moderators can configure AutoMod rule priorities\footnote{\url{https://www.reddit.com/r/reddit.com/wiki/automoderator/full-documentation/}}, but reliance on manually-written syntactic rules makes translating nuanced harm reduction expertise into effective automated actions difficult. As P3 explained, inherited configurations make it ``hard...to completely...understand/make connections'' between keywords and rule violations. When rules accumulate, maintenance becomes unmanageable (P4); complex and outdated rules generate continuous streams of irrelevant flags demanding review, consuming moderators' limited attention and distracting from genuine high-stakes threats. This misallocates expertise: systems require those valued for risk-assessment capabilities to also perform the secondary technical labour of rule-base maintenance. To better support this work, moderation tools must shift from requiring exhaustive low-level rule configuration to enabling high-level expert input.

Large Language Models offer potential mechanisms for this shift through programming-by-example capabilities~\cite{yeh2025bridging}. Instead of writing complex configurations, moderators could provide posts to LLM-powered tools along with high-level commands like ``Generate an AutoMod rule to detect and flag similarly dangerous advice as high priority.'' This reconfigures automated systems from static rule repositories into adaptive partners rapidly taught to recognise and prioritise new harms, allowing expertise to be directly applied to managing risk rather than expended on tool maintenance. However, this introduces risks requiring careful consideration. LLMs produce errors~\cite{wang2025towards} that could cause the broad incorrect removal of valuable harm reduction content. Recent work shows LLMs exhibit inconsistent detection and unstable behaviour under small norm changes~\cite{kolla2024llm, kumar2024watch}, raising concerns about implementation reliability in high-stakes health contexts. Additionally, LLMs trained on public data may encode societal biases stigmatising substance use, potentially amplifying the epistemic injustice moderators work to counteract~\cite{ajmani2024whose}. Responsible implementation must position LLMs as assistants generating rule-based solutions with explanations, keeping expert moderators in final supervisory roles rather than functioning as autonomous decision-makers.

In summary, as global health crises escalate (e.g., mental health and substance use disorder), technologies that harm marginalised communities or ignore the harm they cause become ethically untenable. This research opens questions HCI must address: How do platforms accommodate communities whose health goals conflict with legal liability concerns? What governance models enable expert judgment when that expertise cannot be formally credentialed or its application is legally constrained? How do we design moderation tools that support rather than burden volunteers from marginalised populations? Can annotation schemas capture contested definitions of harm without imposing dominant institutional frameworks? How do communities sustain specialised moderation when recruiting from stigmatised populations with limited resources? 

These questions matter not only for PWUD but for any vulnerable community navigating platform governance that privileges institutional knowledge, legal compliance, and scalable automation over community expertise, health outcomes, and context-sensitive judgment. Answering them requires HCI researchers to work directly with these communities, understand their governance challenges, and develop approaches that centre marginalised voices in platform design rather than treating them as edge cases in systems built for mainstream populations. Our findings extend HCI moderation frameworks~\cite{jiang2023trade} by demonstrating how legal liability structures can systematically undermine protective community work, revealing limitations in governance theories that assume aligned platform-community interests and informing design for contexts where regulatory compliance itself produces harm.

\section{Limitations and Future Work}

Conducting asynchronous interviews via Reddit messaging enabled participation from time-constrained moderators and allowed them sufficient time to reflect on their moderation decisions and reference specific moderation cases. However, we could not pursue spontaneous follow-ups as readily as in synchronous settings, and non-verbal cues were absent. We mitigated this limitation through iterative follow-up questions and by encouraging participants to use emojis and informal language to convey tone. Future research might complement asynchronous methods with observational studies of moderation-in-action to capture real-time decision-making processes.

Our qualitative analysis focused on moderators' experiences from three major PWUD subreddits (r/Drugs, r/MDMA, and r/researchchemicals). While these communities raised a range of concerns and perspectives on moderation challenges, the small sample size and limited contact with participants, limited our ability to explore certain aspects in greater depth, such as differences in practices based on moderator experience or community focus. Future research could examine how moderation practices vary across different platforms, contexts, and community sizes to develop a broader understanding of harm reduction content moderation. 

Interviewing current moderators provided valuable insights into operational challenges but may not fully capture how moderation practices evolve over time or how communities adapt to emerging substance-related risks. Longitudinal studies could better document how PWUD communities develop specialised moderation expertise and respond to changes in platform policies or substance use trends. Additionally, incorporating perspectives from community members could illuminate how moderation decisions affect access to harm reduction information and community participation, thereby better informing moderation practices. The intersection of platform governance and harm reduction efforts revealed in our study points to broader questions about how to better serve vulnerable communities within existing content moderation frameworks, particularly regarding the challenges we identified in this study. Research examining alternative governance models or specialised platform features for harm reduction spaces could help address the tensions that moderators of PWUD communities currently face between community preservation and information access. We urge future work to explore the design of moderation tools that are sensitive to harm reduction contexts, to better accommodate the needs of PWUD for community support while protecting individual safety from potential risks involved in online spaces.

\section{Conclusion}

Our study examined the work of moderating online communities for People Who Use Drugs (PWUD). Our analysis shows that moderators of PWUD communities aim to promote supportive and non-judgmental online spaces to improve the well-being of PWUD. However, their practices constitute a specialised form of public health labour that current sociotechnical systems inadequately support. Moderators in PWUD communities must navigate three distinct challenges that extend existing HCI moderation frameworks~\cite{jiang2023trade}: (1) high-stakes risk evaluation requiring domain-specific expertise, (2) time-critical crisis intervention for members in distress and the monitoring of emergent risks from unregulated drug markets, and (3) the navigation of structural conflicts where platform-wide restrictions around substance use directly oppose their community's harm reduction mission (RQ1). This work reveals how legal liability structures can systematically undermine expert moderators' protective work; this dynamic holds implications for other marginalised communities facing similar regulatory tensions, including abortion care and sex work contexts where platform policies conflict with community health needs.

Within these constraints, moderators are compelled to implement restrictive rules on prohibited content and carefully curate online discussions to enable both community safety and information access (RQ2). They actively foster harm reduction practices among community members and preserve high-quality discussions for community education through sustained engagement. However, current moderation tools, designed for generic rule enforcement, are insufficient for this context. For example, their mechanisms lack the sensitivity to distinguish prohibited substance-sourcing requests from valuable harm reduction information, which limits access to life-saving information by overly restricting community discussion.

This study identifies opportunities for designing sociotechnical systems to better support this essential work (RQ3). We propose two necessary shifts: first, moving from automated decision-making to tools that support human sensemaking in contexts with competing interests through multi-dimensional annotation approaches; and second, shifting from systems requiring low-level rule programming (e.g., manually configuring regular expressions for AutoMod) to those that enable high-level expert guidance (e.g., teaching a tool by providing an example post and instructing it to identify similar content). However, our analysis surfaces unresolved tensions requiring HCI research attention: annotation systems risk overwhelming volunteers already facing sustainability challenges, and incorporating LLMs introduces risks of errors and bias in high-stakes health contexts where decisions affect lives. These insights advance our understanding of how moderation systems can better accommodate PWUD communities while identifying open questions about supporting specialised expertise without overburdening volunteer labour.

% \begin{acks}
% \end{acks}

\bibliographystyle{ACM-Reference-Format}
\bibliography{references}

\end{document}